\documentclass[english]{article}
\usepackage{lmodern}
\usepackage[T1]{fontenc}
\usepackage[latin9]{inputenc}
\usepackage{geometry}
\usepackage{color}
\usepackage{graphicx}

\makeatletter
\newcommand{\lyxaddress}[1]{
\par {\raggedright #1
\vspace{1.4em}
\noindent\par}
}

\makeatother

\usepackage{babel}

\begin{document}

\title{Hints from Tevatron, a prelude to what?}

\author{V. E. \"Ozcan$^{1}$ , S. Sultansoy$^{2,3}$ and G. \"Unel$^{4}$}

\maketitle

\lyxaddress{$^{1}$ Department of Physics and Astronomy, University College London,
London, UK\textcolor{black}{.}\\
\textcolor{black}{$^{2}$ Institute of Physics, Academy of Sciences,
Baku, Azerbaijan. }\\
\textcolor{black}{$^{3}$ TOBB ETU, Physics Department, Ankara, Turkey.}\\
\textcolor{black}{$^{4}$ University of California at Irvine, Physics
Department, USA. }}

\begin{abstract}
We comment on the recent results from the Tevatron experiments in
the W+jets channel and consider some models as the possible underlying
physical theories. We also list some channels for further studies.
\end{abstract}

\section{Introduction}

\begin{figure}
\begin{centering}
\includegraphics[scale=0.7]{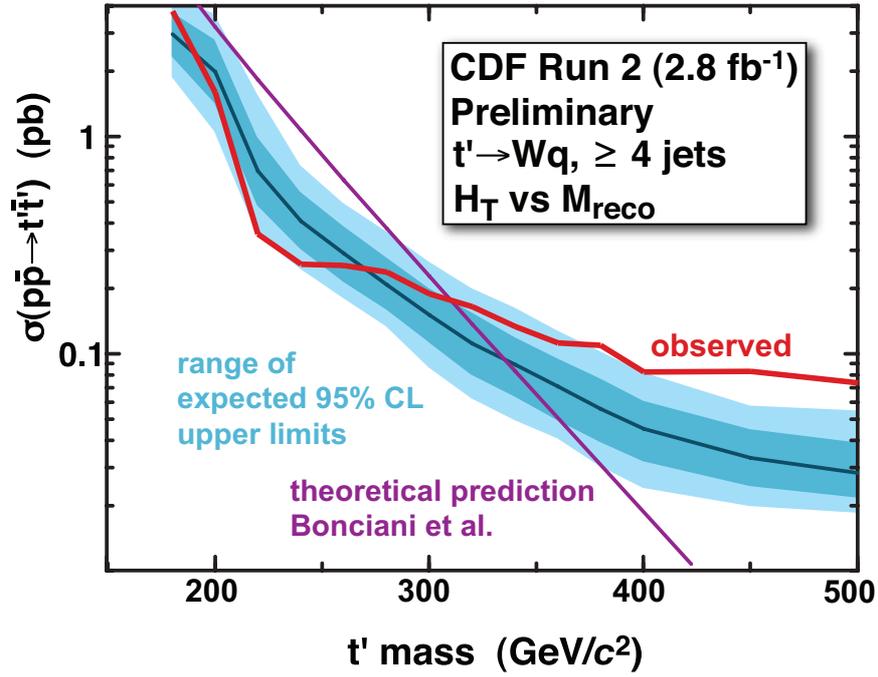}
\par\end{centering}

\caption{Upper limit, at 95\% CL, on the production rate for $t'$ as a function
of $t'$ mass (red) using 2.8 fb$^{-1}$ of integrated luminosity.
The purple curve is the theoretical cross section assuming SM like
couplings of the new quarks. The dark blue band is the range of expected
95\% CL upper limits within one standard deviation. The light blue
band represents two standard deviations .\label{fig:Upper-limit}}

\end{figure}

The Standard Model (SM) is expected to be the low energy limit of
a more fundamental theory \cite{PDG}. The known candidates for such
a theory have more fundamental particles than what is experimentally
known today. Therefore, searches for new particles hence for the new
model of elementary particles and their interactions, continue in
both the precision physics and collider experiments. In a recent public
note, CDF experiment at Tevatron excluded a standard model fourth-generation
$t'$ quark with mass below 311$\,$GeV at 95\% CL using 2.8$\,$fb$^{-1}$
of data (see figure \textcolor{black}{\ref{fig:Upper-limit})}\cite{CDF-public}.
The shown theoretical model shows the tree-level cross section of
a new quark with q=2/3 charge.

The same note also reports an excess of about 5 events in the W+jets
channel in the region between 375-500$\,$GeV. Although this small
number of excessive events can be explained by a detector over-efficiency
or by some unknown SUSY process, in the following text we will consider
some theoretical models where an additional heavy quark is predicted.
Some of these models were also mentioned in the above mentioned note.

\section{Recent CDF measurement on W+jet}

\begin{figure}
\begin{centering}
\includegraphics[scale=0.8]{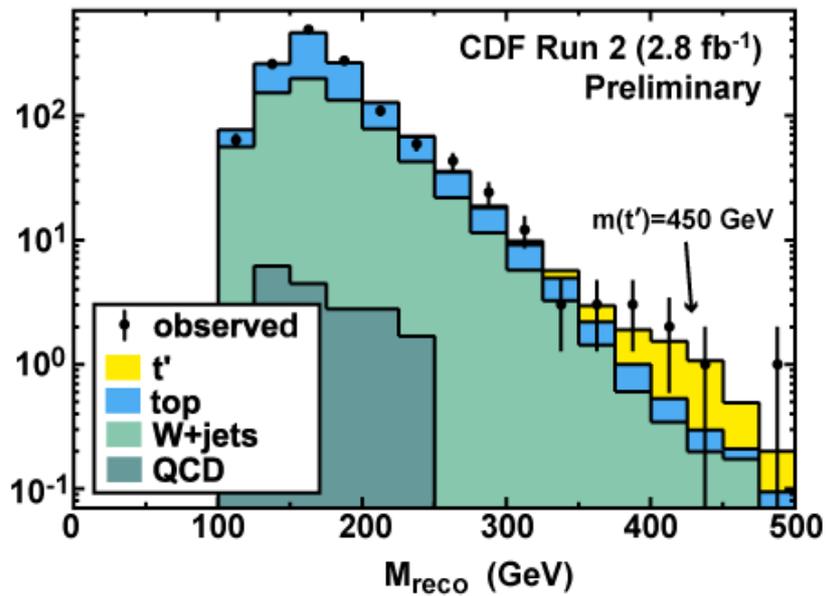}
\par\end{centering}

\caption{The measurement from CDF in the Wjet channel using 2.8 fb$^{-1}$
of integrated luminosity. \label{fig:CDF-measurement}}

\end{figure}

The CDF result on the reconstructed invariant mass in the W+jet channel
is presented in figure \ref{fig:CDF-measurement}. The number of observed
events in the range 375 - 500 GeV is 7 with an expected background
of about 1.8 events. The Poisson probability of such a statistical
deviation is 0.2\%, which is rather low. Taking this excess at its
face value, we calculate its significance, using the well known estimator
\cite{CMS_significance} ${\cal S}=\sqrt{2\times[(s+b)\ln(1+\frac{s}{b})-s]}$,
to be about 2.9$\sigma$, perhaps a hint for a new quark decay. However
the candidate underlying model has to be investigated in the light
of the number of excess events. The model should produce enough cross
section for an excess of 5 events with 2.8 fb$^{-1}$ integrated luminosity.

The efficiency of the CDF event selection can be estimated from the
results in \cite{CDF-public}. The 95\% exclusion limit for $m_{Q}\approx400$
GeV is given as 81 fb. The SM background at 400 GeV can be estimated
using the average of the two adjacent bins, as 0.75 events. Using
the above mentioned signal significance estimator and 0.75 events,
to have 95\% observation limit (i.e. 1.96$\sigma$) one needs 3 signal
events which can be converted to a reconstruction efficiency of $\epsilon=\frac{3.0}{2.8\times81}=0.01$
. The efficiency value can be used to calculate the number of signal
events predicted by the theoretical model shown in figure \ref{fig:Upper-limit},
to see whether it provides enough events to explain the observed excess.
We find that the cross section given in figure \ref{fig:Upper-limit}
can only provide $18\times2.8\times\epsilon$ = 0.5 events. The Poisson
probability of having 2.3 (=$0.5+1.8$) expected events to oscillate
statistically to 7 or more events is 0.9\%. The cross section value
can be increased about 10\% with the QCD scale and PDF uncertainties.
The LO to NLO conversion factor, $k$, for $t\,\bar{t}$ at the Tevatron
is estimated as a number between 1.2 and 1.5 \cite{kfactor}. Folding
in these changes, one can increase the expected number of signal events
to a value of 0.8 events, still quite small to account for the 5 events
in question.

\subsection{Possible Explanations \label{sub:possible-explanations}}

\begin{figure}
\begin{centering}
\includegraphics[scale=0.5]{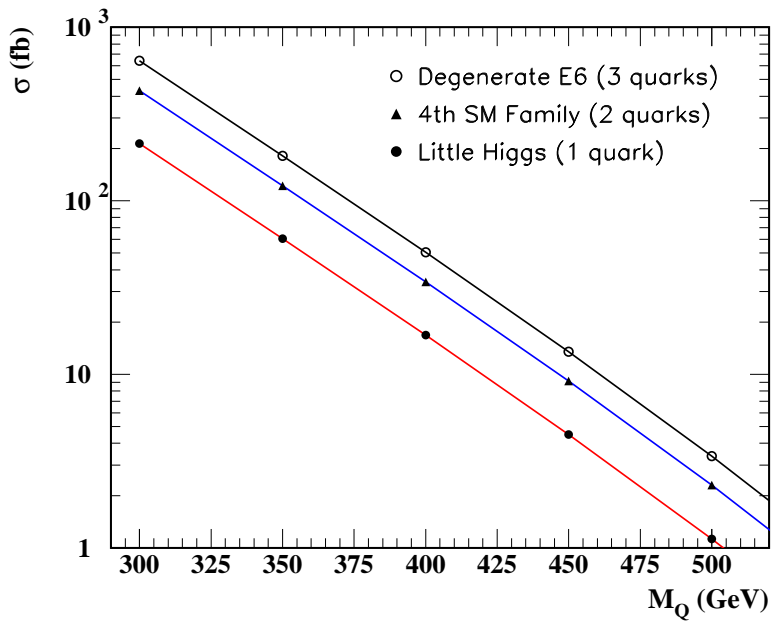}\caption{The expected cross section of the $p\bar{p}\to Q\bar{Q}$ as a function
of the quark mass from Fourth Family, Little Higgs and E6 GUT models
for $\sqrt{s}=1.96$ TeV. In all cases, the PDF set CTEQ6L1 \cite{R-cteq}is
used and the QCD scale is set to the mass of the new quark. \label{fig:expected-cross}}

\par\end{centering}
\end{figure}

As pointed out in the CDF public note \cite{CDF-public}, the fourth
family predicted by the DMM \cite{DMM1,DMM2,DMM3,DMM4,FLDem review}
approach could be the underlying model providing such an excess of
events. The additional quarks in the fourth family are denoted in
the literature as $u_{4}$, $d_{4}$ or $t'$, $b'$ and they are
expected to be quasi-degenerate, i.e. the mass difference between
the up-type and down-type quark to be less than $m_{W}$, disabling
the $u_{4}\to Wd_{4}$ decay channel. Other often cited models with
additional quarks are the E6-GUT\cite{R-e6} and the Little Higgs\cite{LittleHiggs}
models. The latter has only one additional up-type iso-singlet quark,
$T$, whereas the former has one additional down-type iso-singlet
quark per SM family denoted as $D,$ $S$, and $B$ . The case where
the masses of these 3 quarks are close to each other can be considered
as the 'degenerate' E6 model. The figure \ref{fig:expected-cross}
gives the expected cross section of these models as a function of
the (degenerate) quark mass at the Tevatron for QCD scale set to the
mass of the new quark.

If the the forth family quark masses are indeed degenerate, and the
$4\times4$ CKM matrix is such that the fourth family mixings to the
first two families are favoured, the expected number of events can
be calculated as $N=2.8\times\sigma\times k\times\epsilon$, where
$\sigma$ is the tree-level pair production cross section and $k$
=1.5 as given before. For quarks of mass $m_{Q}\approx$ 400$\,$GeV
and with $M_{Q}$ as the QCD scale choice, the cross section is obtained
as 35 fb and $N$ as 1.5 events, whereas for another possible choice
of QCD scale as $2\times m_{W}$, $\sigma$ is found as 57 fb, yielding
N=2.4 events. In these two scenarios, the expected number of events
become 3.3 and 4.2, respectively, bringing in better compatibility
with the measurement of 7 events. The error on these event yields
originating from the PDF choice is about 7\%. A more favorable event
yield might be obtained if the mass difference between the up-type
and down-type quarks is assumed to be slightly larger but still smaller
than $m_{W}$. For example, if $m_{t'}$ =375 GeV and $m_{b'}$ =
400 GeV, the expected number of total events become 3.9 and 5.3 for
the two considered QCD scales. In this case, the probability of measuring
7 events or more becomes 10\% and 28\% respectively.

Such an excess could also arise from the E6 GUT models when the additional
iso-singlet object in the 27-plet, has quark-like properties. In the
case of an E6 GUT with degenerate iso-singlet quark mass values of
approximately 400 GeV, the pair production cross section at the tree
level is 50.4 fb for QCD scale of $m_{Q}$ and 84.9 fb for QCD scale
of $2\times m_{W}$. The branching fraction of $D\to Wj$ is 67\%
and reduces the number of the expected heavy quark CC decay events
to the level of the above discussed fourth family case: 1.5 and 2.5
signal events. However if two of the quarks have masses around 375
GeV and one around 400 GeV, one finds 5.6 expected events for the
measurement of 7 events yielding an occurrence probability over 31\%.

The Little Higgs model also predicts an additional quark whose the
production cross section is one-third of the 'degenerate' E6 GUT model
at the Tevatron. However, as this model has only one additional quark
it is unlikely to provide enough cross section in a trivial way.

\subsection{A cross check with the Higgs searches in CDF and D0}

\begin{figure}
\begin{centering}
\includegraphics[scale=0.4]{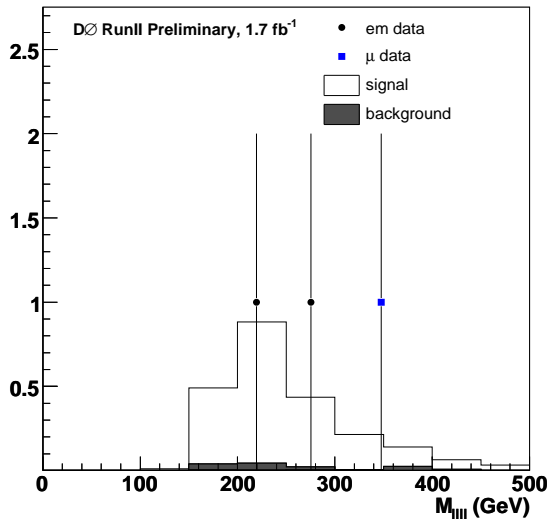}
\par\end{centering}

\caption{The D0 measurement on $p\bar{p}\to ZZ\to4\ell$ where $\ell$ can
be an electron or a muon. }
\label{fig:D0-higgs-plot}

\end{figure}

Recently both CDF and D0 reported the results of their searches for
$ZZ$ production in four-lepton channel \cite{ZZ}. The most recent
D0 measurement is given in figure \ref{fig:D0-higgs-plot} for 1.7
fb$^{-1}$ integrated luminosity. The SM $p\bar{p}\to ZZ\to4\ell$
cross section is experimentally measured as 7.9 fb. The D0 data is
with 1.7 fb$^{-1}$ integrated luminosity, resulting in 13.4 expected
events before event selection and reconstruction cuts. The integration
of the shaded area in the same figure gives 2.4 events yielding a
reconstruction efficiency $\epsilon$ of 17.9\%. If the previously
discussed excess is really coming from one or more new quarks, their
influence on other processes, such as Higgs boson production, should
be cross checked. The new quarks in the mass region 375 - 500 GeV
will strongly influence the Higgs production at the Tevatron, thus
possibly manifesting themselves in the $ZZ$ channel as well. For
example, the fourth family quarks would lead to substantial enhancement
of $ggH$ cross-section and thus would give a chance for Tevatron
to observe the SM Higgs. For a Higgs of mass 200 GeV, the effective
cross section of 4 lepton events was calculated to be 1.5 fb in the
presence of a fourth SM generation\cite{ggh_FF}. In this case, the
expected number of $H \to ZZ \to4\ell$ events is $1.5\times1.7\times0.179=0.46$,
making the agreement between the calculations and data even better:
$2.4+0.46\approx2.9$ expected events versus 3 observed events. 

The iso-singlet quarks might also boost the Higgs searches via their
possible NC decays, i.e. using the $D\to H\, jet$ channel\cite{E6-higgs}.
However even if the total luminosity is tripled, the Tevatron will
not be able to benefit from this mode, one will have to wait for the
LHC data.

In a recent Fermilab press-release, D0 results were announced as the
\textquotedblleft{}Prelude to the Higgs\textquotedblright{} keeping
in mind the $ZH$ production \cite{pressRelease}. However, in $ZH$
channel, with 2.3 fb$^{-1}$ integrated luminosity, the ratio of observed
limit to the SM prediction still exceeds 10. Therefore, this channel
is quite unlikely to be seen before 2010, even when CDF and D0 data
are combined. As a second option one could consider \textquotedblleft{}golden\textquotedblright{}
mode keeping in mind the above-mentioned enhancement in four SM family
case. The Higgs mass regions around 150 GeV and 200-240 GeV could
be reached by the combined data at the end of 2009. Unfortunately,
both opportunities seem to be rather pessimistic.

\section{Some suggestions \label{sec:Suggested-channels}}

In section \ref{sub:possible-explanations}, we estimated the same
number events for both the fourth family and E6 GUT models. The method
to distinguish between these two would be a search for the NC decays,
namely the $t'\to Z+jet$ channel. Since there is no FCNC at the tree
level in the models with fourth SM family, the presence of a signal
in this channel would motivate preference of E6 GUT models over the
fourth family models. However as the branching fraction of $D\to Z+jet$
is 33\%, the expected excess at the same integrated luminosity of
2.8 fb$^{-1}$ would be 2-3 events depending on the QCD scale if the
same event selection and reconstruction efficiency is assumed. A suggested
mode would be $p\bar{p}\to t'\bar{t'}\to WjZj$ where leptonic decays
of the $W$ branch could be used for trigger and the $Z$ branch for
the reconstruction purposes. In order to increase the statistics the
combination of results between the two Tevatron experiments is obligatory.

If the four family SM is realized in Nature, then the Higgs boson
could be searched in the so-called \textquotedblleft{}silver\textquotedblright{}
mode, where the Higgs decays to two heavy neutrinos \cite{silver_mode}.
If both the Higgs boson and the fourth family neutrino have appropriate
masses, then the $p\bar{p}\to h\to\nu\nu\to2W\,2\mu$$ $ chain would
be the channel to look for the Higgs boson, especially if $\nu$ is
of Majorana nature, thus providing same sign leptons in the final
state. For example if $m_{H}$=200 GeV and $m_{\nu}$=90 GeV, the
$\sigma(p\bar{p}\to h)=1337$ fb and BR($H\to\nu\nu$ ) is about 6\%,
resulting in 80 fb effective cross section. Taking the 4x4 extension
of MNS matrix compatible with the neutrino mixing measurements \cite{MNS-4x4},
the BR($\nu\to W\mu$ ) is found as 0.68. Using the 2.8 fb$^{-1}$
integrated luminosity reported in \cite{CDF-public}, this channel
yields more than 100 $WW\mu\mu$ events out of which 50 would be with
same sign leptons.

\section{Conclusions}

In the light of the recent data from the Tevatron, there might already
be a hint for beyond the 3 family SM physics, yet to be discovered.
We propose degenerate fourth family and E6-GUT models with quarks
around 400 GeV to explain the observed excess of events. Clearly,
more data from the Tevatron is needed to understand the situation
completely, as well as the study of the channels suggested in section
\ref{sec:Suggested-channels} by both collaborations together with
the combination of their experimental results. On the other hand,
we believe this ``hint'' is a prelude to the BSM physics to come
from the LHC at the early stages of the data taking \cite{Q-LHC-review}.
For example, degenerate fourth family quarks with mass 400 GeV would
be discovered at 5$\sigma$ level with 80pb$^{-1}$ integrated luminosity
\cite{FF-erkcan}, i.e. during the first weeks. Similarly, an iso-singlet
$D$ quark with 400 GeV mass would be discovered with 1fb$^{-1}$,
i.e. during first year \cite{E6-pub1}. Or, 300 GeV mass Higgs boson
which corresponds to quartic Higgs boson coupling constant equal to
$g_{w}$ will be discovered with 200pb$^{-1}$ (4fb$^{-1}$) in 4
(3) SM family case \cite{ff_enhancement}. Therefore, a rapid startup
and commissioning of the LHC beams is needed to provide enough luminosity
to both general-purpose detectors on this new energy frontier. 

Finally, it seems that the next two years will be very competitive.

\subsection*{Acknowledgments}

S.S. acknowledges the support from the Turkish State Planning Committee
(DPT) and Turkish Atomic Energy Authority (TAEK). G.U.'s work is supported
in part by U.S. Department of Energy Grant DE FG0291ER40679. V.E.O.
acknowledges financial support from the UK Science and Technology
Facilities Council.

\end{document}